\pdfoutput=1

\newcommand*{\ATLASLATEXPATH}{}

\documentclass[cernpreprint,texlive=2016,txfonts,UKenglish,texmf]{\ATLASLATEXPATH atlasdoc}

\usepackage[subfigure]{\ATLASLATEXPATH atlaspackage}

\usepackage{\ATLASLATEXPATH atlasbiblatex}

\usepackage{\ATLASLATEXPATH atlascontribute}

\usepackage{\ATLASLATEXPATH atlasphysics}

\graphicspath{}

\usepackage{PaperDraft-defs}
\usepackage{multirow}
\usepackage{epstopdf}
\usepackage{textcomp}
\usepackage{siunitx}
\usepackage{booktabs}

\AtlasTitle{Search for new phenomena in dijet events using 37~\ifb\ of $pp$ collision data collected at $\sqrt{s}=13$~\TeV\ with the ATLAS detector}

\author{The ATLAS Collaboration}

\AtlasRefCode{EXOT-2016-21}

\AtlasNote{ATL-COM-PHYS-2016-1495}

\PreprintIdNumber{CERN-EP-2017-042}

\AtlasDate{\today}

\AtlasJournal{PRD}

\AtlasAbstract{%
Dijet events are studied in the proton--proton collision dataset recorded at $\sqrt{s}=13$~\TeV\ with the ATLAS detector at the Large Hadron Collider in 2015 and 2016, corresponding to integrated luminosities of 3.5~\ifb\ and 33.5~\ifb\ respectively. Invariant mass and angular distributions are compared to background predictions and no significant deviation is observed. For resonance searches, a new method for fitting the background component of the invariant mass distribution is employed. The dataset is then used to set upper limits at a 95\% confidence level on a range of new physics scenarios. Excited quarks with masses below~\qstarLimitMjj\ are excluded, and limits are set on quantum black holes, heavy \Wprime\ bosons, \Wstar\ bosons, and a range of masses and couplings in a \Zprime\ dark matter mediator model. Model-independent limits on signals with a Gaussian shape are also set, using a new approach allowing factorization of physics and detector effects. From the angular distributions, a scale of new physics in contact interaction models is excluded for scenarios with either constructive or destructive interference. These results represent a substantial improvement over those obtained previously with lower integrated luminosity.
}

\hypersetup{pdftitle={Search for new phenomena in dijet events with the ATLAS detector at sqrt(s)=13 TeV using the full 2015 and 2016 datasets},pdfauthor={The ATLAS Collaboration}}

\begin{document}

\maketitle

\section{Introduction}
\label{sec:intro}

The Large Hadron Collider (LHC)~\cite{Evans:2008zzb} at CERN has been colliding protons at a center-of-mass energy of $\sqrt{s}=13$~\TeV\xspace since 2015. With the completion of the 2016 physics run, the total integrated luminosity of Run-2 data at $13$~\TeV\ now exceeds that of the total Run-1 dataset by more than 10~\ifb. When combined with the increase in parton luminosity~\cite{Campbell:2006wx} at high energy scales, due to the raising of the center-of-mass energy from $8$~\TeV\xspace to $13$~\TeV, this very large dataset provides an exceptional opportunity to search for new phenomena.

New particles directly produced in proton--proton ($pp$) collisions must interact with the constituent partons of the proton and, consequently, can produce partons when they decay.
Such partonic final states dominate in many models of new phenomena beyond the Standard Model (BSM) which are accessible at the LHC.
The partons shower and hadronize, creating collimated jets of particles carrying approximately the four-momenta of the partons.
The production rates for BSM signals decaying to two-jet (dijet) final states can be large, allowing such signals to be probed through searches for anomalous dijet production at masses constituting significant fractions of the total hadron collision energy.

In the Standard Model (SM), hadronic collision production of jet pairs primarily results from $2\rightarrow 2$ parton scattering processes via strong interactions described by quantum chromodynamics (QCD).  Particles emerge from these collisions as jets with high transverse momentum (\pt) with respect to the incoming partons.
A smooth and monotonically decreasing distribution for the dijet invariant mass, \mjj, is predicted by QCD~\cite{Harris:2011bh}.
The presence of a new resonant state decaying to two jets may introduce an excess in this distribution, localized near the mass of this resonance.
Furthermore, in QCD most dijet production occurs in the forward direction at small angles $\theta^*$, defined as the polar angle with respect to the direction of the initial partons in the dijet center-of-mass frame,\footnote{Since, experimentally, the two partons cannot be distinguished, $\theta^*$ is always taken between 0 and $\pi/2$.} due to $t$-channel poles in the cross-sections for the dominant scattering processes.
Many theories of BSM physics predict additional dijet production with a more isotropic signature, and thus a significant population of jets produced at large $\theta^*$~\cite{Harris:2011bh,BoelaertAkesson:2010}.
The search reported in this \publicationType exploits these generic features of BSM signals in an analysis of the dijet mass and angular distributions. Following a model-nonspecific search for deviations from the SM in both types of distributions, limits are set on the masses of excited quarks, quantum black holes, \Wprime\ and \Zprime\ bosons, and excited chiral \Wstar\ bosons, on contact interactions scales, and on generic Gaussian-shaped signal production.

Results from prior investigations of dijet distributions with lower-energy hadron collisions at the S$\mathrm{\bar{p}}$pS~\cite{Arnison:1983dk,Albajar:1988rs,Bagnaia1984283}, 
the Tevatron~\cite{CDF:2009DijetSearch,DZero:2009DijetAng}, and the 
LHC at $\sqrt{s}=7$--8~\TeV\xspace~\cite{EXOT-2010-01,EXOT-2010-02,CMS-EXO-10-010,CMS-EXO-10-002,CMS-QCD-10-016,CMS-EXO-11-015,EXOT-2010-07,EXOT-2011-07,EXOT-2011-21,EXOT-2013-11,CMS-EXO-12-016,EXOT-2014-15} 
were found to be in agreement with QCD predictions.  Recent searches at 13 TeV~\cite{CMS-EXO-15-001,EXOT-2015-02,CMS-EXO-16-032}
included extensions of the analysis to di-$b$-jet final states~\cite{EXOT-2015-22}
and to lower masses~\cite{CMS-EXO-16-032,CMS-EXO-14-005}, 
and observed no significant deviations from the Standard Model.
This \publicationType\ presents an analysis of the full 2015 and 2016 datasets recorded by the ATLAS detector at the LHC, corresponding to \integLumi of $pp$ collision data at $\sqrt{s}=13$~\TeV.

\section{ATLAS detector}
\label{sec:detector}

\newcommand{\AtlasCoordFootnote}{
	ATLAS uses a right-handed coordinate system with its origin at the nominal interaction point (IP) in the center of the detector and the $z$-axis along the beam line.
	The $x$-axis points from the IP to the center of the LHC ring, and the $y$-axis points upwards.
	Cylindrical coordinates $(r,\phi)$ are used in the transverse plane, $\phi$ being the azimuthal angle around the $z$-axis.
	The pseudorapidity is defined in terms of the polar angle $\theta$ as $\eta = -\ln \tan(\theta/2)$.
	It is equivalent to the rapidity for massless particles.
}

The ATLAS experiment~\cite{PERF-2007-01,ATLAS-TDR-19} at the LHC is a multipurpose particle detector with a forward-backward symmetric cylindrical geometry with layers of tracking, calorimeter, and muon detectors over nearly the entire solid angle around the $pp$ collision point.\footnote{\AtlasCoordFootnote\xspace}
The directions and energies of high-\pT{} hadronic jets are measured using silicon tracking detectors and a transition radiation straw-tube tracker, hadronic and electromagnetic calorimeters, and a muon spectrometer.
Hadronic energy measurements are provided by a calorimeter with scintillator active layers and steel absorber material for the pseudorapidity range $|\eta| <  1.7$, while electromagnetic (EM) energy measurements are provided by a calorimeter with liquid argon (LAr) active material and lead absorber material covering the pseudorapidity range $|\eta|< 3.2$.
The endcap and forward regions, extending up to $|\eta| = 4.9$, are instrumented with LAr calorimeters for both EM and hadronic energy measurements.
The lower-level trigger is implemented in hardware and uses a subset of the detector information to reduce the accepted rate to \SI{100}{\kilo\hertz}.
This is followed by a software-based high-level trigger that reduces the rate of events recorded to \SI{1}{\kilo\hertz}~\cite{Aaboud:2016leb}.

\section{Event selection}
\label{sec:selection}

Groups of contiguous calorimeter cells (topological clusters) are formed based on the significance of local energy deposits over calorimeter noise~\cite{ATLAS-LARG-PUB-2008-002,PERF-2014-07}. 
Topological clusters are grouped into jets using the anti-$k_t$ algorithm~\cite{antikT,Cacciari:2006} with radius parameter $R=0.4$. 
Jet four-momenta are computed by summing over the topological clusters that constitute each jet, treating the energy of each cluster as resulting from a 
four-momentum with zero mass. 
Jets with \pT~above 20 GeV are reconstructed with an efficiency of nearly 100\%.
Jet calibrations derived from simulation are used to correct the jet energies and directions to those of particle-level jets from the hard-scatter interaction clustered with the same algorithm and parameters\footnote{
The `particle level' jets are built from stable particles defined by having a proper mean decay length of $c\tau>10$ mm.
Particles from interactions other than the hard scattering, as well as muons and neutrinos, are not included in this definition.
More information about the particle definition can be found in Ref.~\cite{ATLAS-PHYS-PUB-2015-013}.
}.
This calibration procedure~\cite{PERF-2011-03,PERF-2012-01,PERF-2014-03, ATLAS-CONF-2015-002,Alkire:2018215}, is followed by a residual calibration accounting for the differences between data and simulation, beginning with a correction to the relative response for forward jets ($|\eta| > 0.8$) with respect to central jets ($|\eta| < 0.8$).
Using this method and other \emph{in situ} techniques where a jet to be calibrated is balanced against a well-calibrated reference object~\cite{ATLAS-CONF-2015-017,ATLAS-CONF-2015-057}, analysis of jet data at 13~\TeV~corrects the jet response and contributes to the uncertainty estimates up to jet \pT values of $2.3$~TeV, beyond which the calibration is frozen.

The total jet energy scale uncertainty is 1\% for central jets with \pT~of 500~\GeV\ and grows to 3\% for jets with \pT~of 2~\TeV, at which point, due to the limited size of the event sample available for the \emph{in situ} studies, an uncertainty is derived from alternative methods using the single-particle response measurements described in Ref.~\cite{PERF-2015-05}.
Uncertainty in the jet energy resolution has a negligible impact on the analysis.
The dijet mass resolution is 2.4\% and 2.0\% for dijet masses of 2 \TeV and 5 \TeV, respectively, derived at 13~\TeV\ from the simulation of QCD processes as in Ref.~\cite{EXOT-2015-02}. 

Collision events are recorded using a trigger that requires at least one jet reconstructed by the high-level trigger
with a \pT greater than \trigJetMin~\GeV, the lowest-\pT single-jet trigger that saves all events that activate it.
Events containing at least two jets are selected for offline analysis if the \pT of the leading (subleading) jet is greater than 440~(60)~\GeV.
This requirement ensures a trigger efficiency of at least 99.5\% for collisions that enter into the analysis.
Events are discarded from the search if any jets with $\pT>$~60~\GeV~are compatible 
with non-collision background or calorimeter noise \cite{ATLAS-CONF-2015-029}. 

\section{Monte Carlo simulation}
\label{sec:simulation}

Monte Carlo (MC) events from multijet production described by QCD are generated with \textsc{Pythia~8.186}~\cite{pythia8} using the A14~\cite{A14tune} set of tuned parameters for the underlying event and the leading-order NNPDF2.3~\cite{Ball:2012cx} parton distribution functions (PDFs).
The renormalization and factorization scales are set to the average \pT~of the two leading jets.
Detector effects are simulated using \textsc{Geant4}~\cite{Agostinelli:2002hh} within the ATLAS software infrastructure~\cite{Aad:2010ah}.
The same software used to reconstruct data is also used to reconstruct simulated events.
The simulated events are used to provide a background estimate for the dijet angular distributions, to test the data-based background estimate used for the \mjj\ distribution, and to provide qualitative comparisons to kinematic distributions in data.

\textsc{Pythia} calculations use matrix elements that are at leading order in the QCD coupling constant, with simulation of higher-order contributions partially covered by the parton shower modeling. 
They also include modeling of hadronization effects.
The distributions of events predicted by \textsc{Pythia} are reweighted to next-to-leading-order~(NLO) predictions of \textsc{NLOJET}++~\cite{NLOJETPP1, NLOJETPP2, NLOJETPP3} using mass- and angle-dependent correction factors defined as in Ref.~\cite{EXOT-2014-15}.
The correction factors modify the shape of the angular distributions at the level of 15\% at high values of \mjj\ and low rapidity separation between the leading and subleading jets. The correction is 5\% or less for the highest values of rapidity separation. 
The \textsc{Pythia} predictions also omit electroweak effects. 
These are included as additional mass- and angle-dependent correction factors~\cite{Sudakov} 
that differ from unity by up to 3\% in the $\mjj > 3.4$~\TeV~region.
The \textsc{Pythia} distributions corrected for NLO and electroweak effects are compared to the angular and \mjj\ distributions in data and are found to be in good agreement within experimental uncertainties.

Signal samples are generated as described in Section~\ref{sec:signals} for a range of benchmark models: excited quarks (\qstar)~\cite{Baur:1987ga,Baur:1989kv}, new heavy vector bosons (\Wprime, \Zprime)~\cite{Altarelli:1989ff,LHCDMF:2015,Chala:2015ama}, excited chiral bosons (\Wstar)~\cite{Chizhov:2010jg,Chizhov:2009fc}, quantum black holes (QBH)~\cite{Gingrich:2009hj,Calmet:2008dg,Dai:2007ki} and contact interactions (CI)~\cite{CI1, CI3}.
 After these signals are simulated, most of the samples are reconstructed using the same framework as used for QCD processes, though a small fraction of the samples employ a simplified parameterization of the detector as described in Ref.~\cite{ATL-SOFT-PUB-2014-01} for improved processing time. No difference between full simulation and this fast simulation is observed in the relevant variables for this analysis.

\section{Resonance search}
\label{sec:resonant} 

\begin{figure}[h!]
  \centering
  \subfigure[$|\yStar| < 0.6$ selection]{\includegraphics[width=0.49\columnwidth]{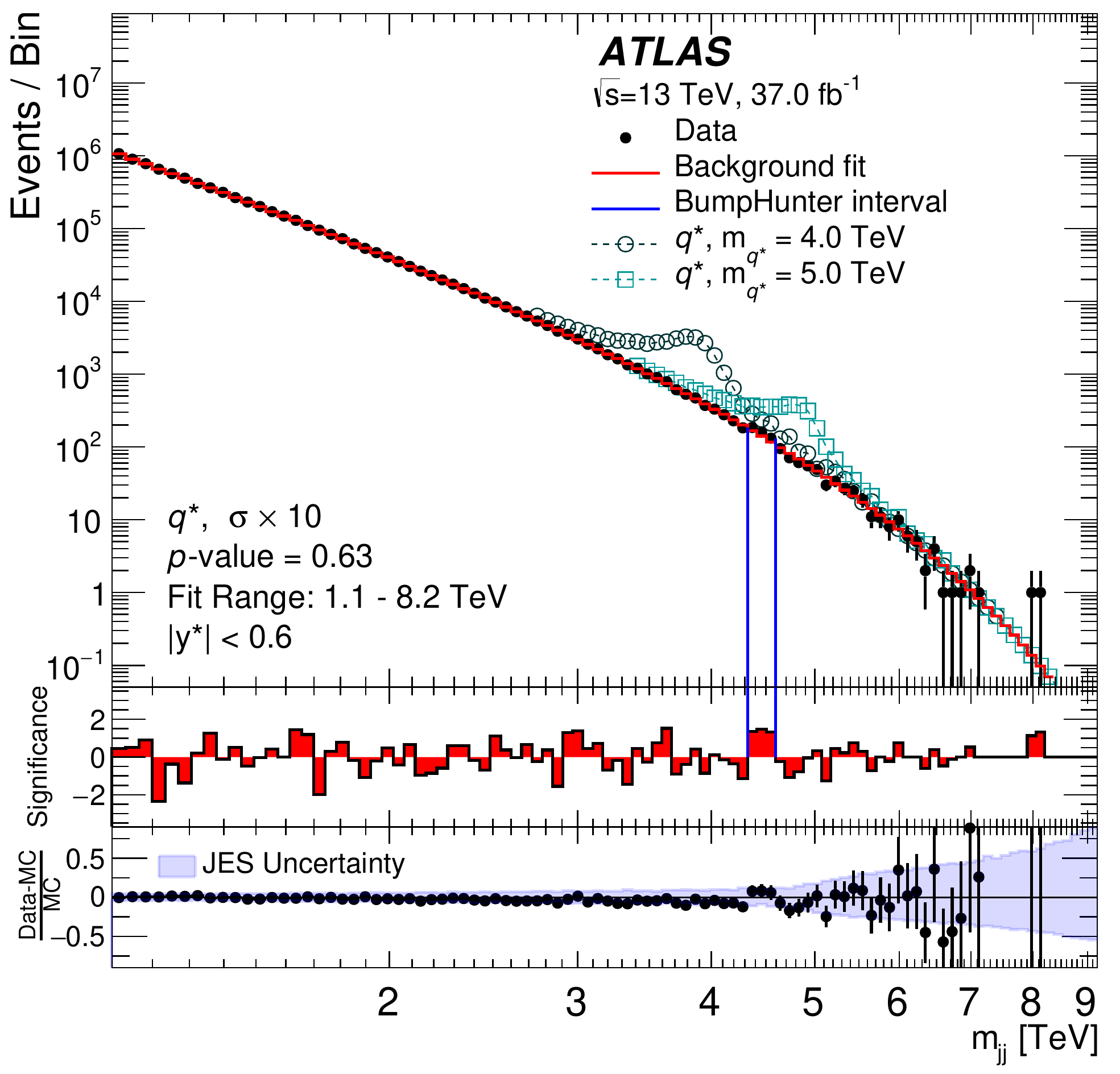}\label{subfig:resSearch}}  
  \subfigure[$|\yStar| < 1.2$ selection]{\includegraphics[width=0.49\columnwidth]{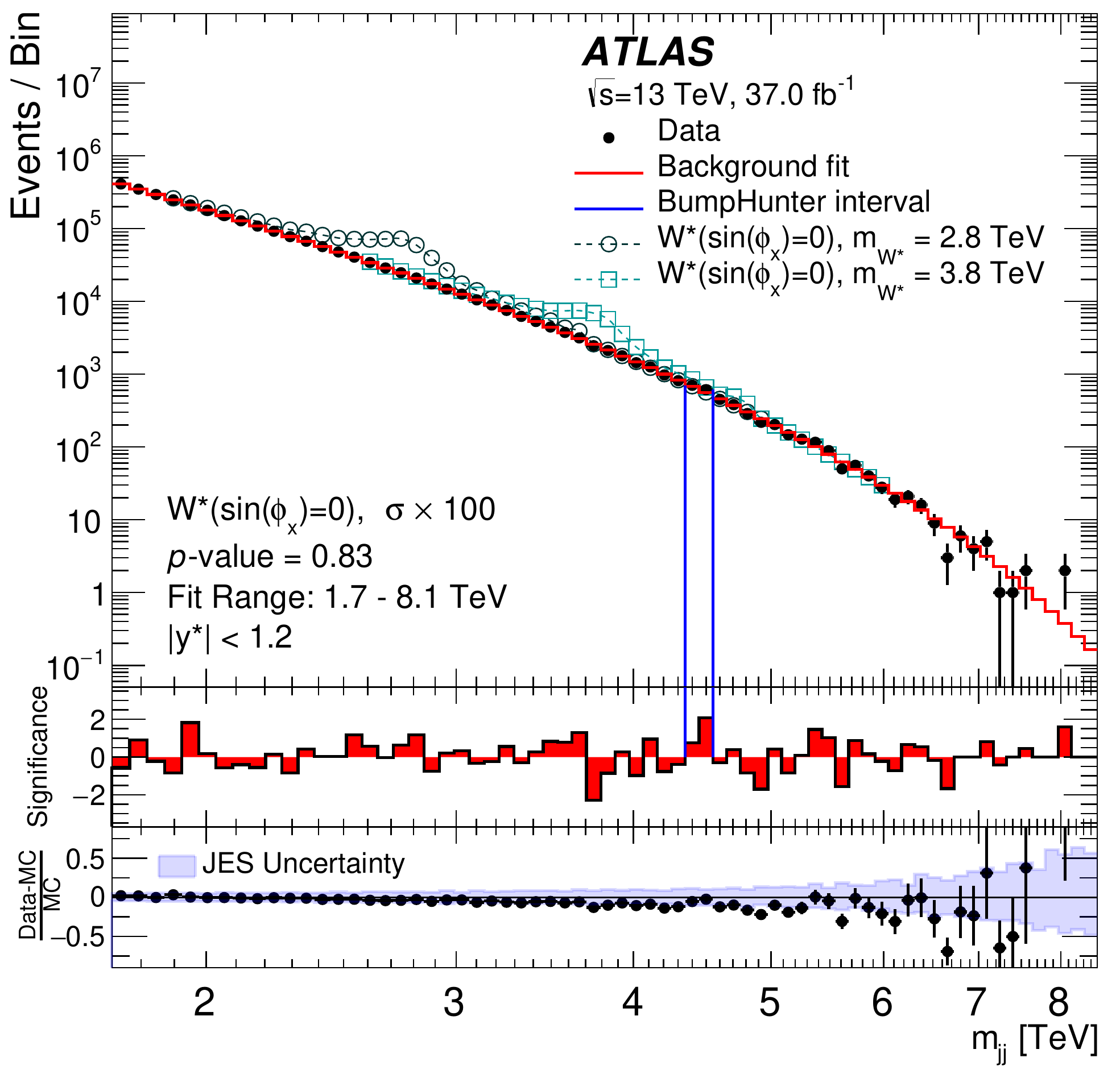}\label{subfig:angSearch}} 
  \caption{The reconstructed dijet mass distribution $\mjj$ (filled points) is shown for events with 
    $\pT>440$~$(60)$~$\GeV\xspace$ for the leading~(subleading) jet. The spectrum with $|y^*|~<~0.6$ is shown in (a) for events above $\mjj = 1.1$~\TeV\ while the selection with $|y^*|~<~1.2$ is shown in (b) for events above $\mjj = 1.7$~\TeV.
    The solid line depicts the background prediction from the sliding-window fit.
    Predictions for benchmark signals are normalized to a cross-section large enough to make the shapes distinguishable above the data.
    The vertical lines indicate the most discrepant interval identified by the \textsc{BumpHunter} algorithm, for which the $p$-value is stated in the figure. 
    The middle panel shows the bin-by-bin significances of the data--fit differences, considering only statistical uncertainties.
    The lower panel shows the relative differences between the data and the prediction of \textsc{Pythia~8} simulation of QCD processes, corrected for NLO and electroweak effects, and is shown purely for comparison.  The shaded band denotes the experimental uncertainty in the jet energy scale calibration.
    }
  \label{fig:one}
\end{figure}

The \mjj distribution formed from the two leading jets in selected events is analyzed for evidence of contributions from resonant BSM phenomena. The rapidity of an outgoing parton is $y =$~\textonehalf~$\ln{[(E+p_z)/(E-p_z)]}$,
where $E$ is its energy and $p_z$ is the component of its momentum along the $z$-axis. The rapidity difference $\yStar = (y_1-y_2)/2$ is defined between the two leading jets and is invariant under Lorentz boosts along the $z$-axis.  A requirement of $|\yStar| < 0.6$ reduces the background from QCD processes. 
This nominal selection is used for the model-independent search phase, to set limits on generically-shaped signals (discussed in Section~\ref{sec:signals}), and to constrain the \qstar, QBH, \Wprime\, and \Zprime\ benchmark models, all of whose distributions peak at $\yStar = 0$. 
A second signal region with a wider selection of $|\yStar| < 1.2$ is also defined, optimized for signals produced at more forward angles.
The \Wstar\ benchmark model, whose distribution peaks at $|\yStar| > 1.0$, is constrained using this selection.
Due to the requirements on \yStar\ and \pt\, the selection is fully efficient only for $\mjj> 1.1$~\TeV\ (1.7 TeV for the $|\yStar| < 1.2$ selection). Therefore, the analysis is performed above this mass threshold. Bin widths are chosen to approximate the \mjj resolution and therefore widen as the mass increases, from about 130 GeV at the lowest \mjj\ values to about 180 GeV at the highest. They differ slightly between the $|\yStar| < 0.6$ and $|\yStar| < 1.2$ selections as the resolution also differs.

Figure \ref{fig:one} shows the observed \mjj distribution for events passing the two $\yStar$ selections, overlaid with examples of the signals described in Section~\ref{sec:signals}. The background estimate is illustrated by the solid red line and is derived from the sliding-window fitting method described below. The largest value of \mjj detected is \HighestDijetMassResonance.

Prior dijet searches found that expressions of the form:
\begin{equation}
\label{eq:backgroundfit}
f(z)=p_{1}(1-z)^{p_{2}}z^{p_{3}}z^{p_4 \log{z}}, 
\end{equation}
where $z = m_{jj}/\sqrt{s}$ and the $p_{i}$ are parameters, describe dijet mass distributions observed at lower collision energies.  Some past searches required fewer terms in Eq.~\ref{eq:backgroundfit}, such as by setting $p_4=0$, but more parameters are ultimately required to describe the distribution as integrated luminosity increases~\cite{EXOT-2015-02}. Searches at CDF, as well as at ATLAS and CMS at both $\sqrt{s}=8$~and $\sqrt{s}=13$~\TeV, previously found Eq.~(\ref{eq:backgroundfit}) to fit the observed spectrum~\cite{CDF:2009DijetSearch,EXOT-2010-01,CMS-EXO-11-015,EXOT-2010-07,CMS-EXO-16-032,EXOT-2013-11}. This parameterization also provides a good description of simulated QCD samples.

With increasing luminosity and the corresponding extension of the \mjj range and decrease in statistical uncertainties, a single global fit to the entire spectrum using Eq.~(\ref{eq:backgroundfit}) cannot necessarily be relied upon. Since the global fit is still viable for this analysis, it presented an opportunity to develop new methods for addressing the background estimate. For the resonance search in this paper, a new sliding-window fitting technique is used, fitting only restricted regions of the spectrum and therefore retaining more flexibility. The limited range of the sliding-window fit allows the use of a three-parameter fit function, while the global fit requires a non-zero $p_4$. The sliding-window fit produces search and limit results compatible with those from the global fit used in previous analyses. The reliability of this new background fitting method in presence of a signal has also been checked. Tests performed for the full range of signal widths considered in this paper have shown good linearity between the injected and extracted signal. 

The background for the invariant mass spectrum is constructed bin-by-bin by performing a likelihood fit to the data in each window and using the fit value at the center of the window for the background description. The values from the full set of windows are then joined to create the background for the full mass range. The window size is chosen such that the three-parameter version of Eq.~(\ref{eq:backgroundfit}) describes the data well in each window of the fit. The nominal window size covers approximately half of the total number of bins seen in Figure \ref{fig:one}, wide enough for all the considered benchmark signals to fit within an individual window.

The uncertainty due to the values of the parameters in Eq.~(\ref{eq:backgroundfit}) is estimated by repeating the sliding-window fitting procedure on pseudo-data drawn via Poisson fluctuations from the nominal background prediction, that is, the fit result in data.
The uncertainty in each \mjj bin is taken to be the root mean square of the fit results for all pseudo-experiments in that bin.
To estimate an uncertainty due to the choice of background parameterization, an additional sliding-window fit using Eq.~(\ref{eq:backgroundfit}) with $p_4 \neq 0$ is compared to the nominal ansatz, and the average difference between the two fit results across a set of pseudo-data is taken as an uncertainty.
This background prediction for the \mjj distribution does not involve simulated collisions and is therefore
not affected by uncertainties such as those due to MC modeling and statistics.

The \textsc{BumpHunter} algorithm quantifies the statistical significance of any localized excess in the \mjj distribution~\cite{Aaltonen:2008vt, Choudalakis:2011bh}.
The algorithm compares the binned \mjj distribution of the data to the fitted background estimate, considering contiguous mass intervals in all possible locations, from a width of two bins to a width of half of the distribution.
For each interval in the scan, it computes the significance of any excess found.
The algorithm identifies the interval \bumpHunterSignalInterval, indicated by the two vertical lines in Figure~\ref{fig:one}, as the most discrepant interval in the $|\yStar| < 0.6$ signal region.
The global significance of this outcome is evaluated using the ensemble of possible outcomes across all intervals scanned, by applying the algorithm to pseudo-data samples drawn randomly from the background fit. 
Without including systematic uncertainties, the probability that fluctuations of the background model would produce an excess at least as significant as the one observed in the data anywhere in the distribution (the \textsc{BumpHunter} probability) is \bumpHunterPValue.
Thus, there is no evidence of a localized contribution to the mass distribution from BSM phenomena. 
Similarly, the search in the second signal region with $|\yStar| < 1.2$ shows no significant deviation from the smooth background parameterization, with the same interval identified as the most discrepant and a \textsc{BumpHunter} probability of \bumpHunterPValueWStar.

\section{Angular analysis}
\label{sec:angular}

Differences between the rapidities of two jets are invariant under Lorentz boosts along the $z$-axis, hence the following function of the rapidity difference $\yStar$,

\begin{equation*}
\label{eq:chidefinition}
\chi = {\mathrm{e}}^{2|\yStar|} \sim \frac{1+\cos{\thetaStar}}{1-\cos{\thetaStar}} ,
\end{equation*}

\noindent is the same in the detector frame as in the partonic center-of-mass frame.  The variable $\chi$ is constructed such that, in the limit of massless parton scattering and when only $t$-channel scattering contributes to the partonic cross-section, the angular distribution $dN/d\chi$ is approximately independent of $\chi$~\cite{Nele:tidied}.

In the center-of-mass frame, the two partons have rapidity $\pm \yStar$. A momentum imbalance between the two incident partons boosts the center-of-mass frame of the collision with respect to the laboratory frame along the $z$ direction by

\begin{equation*}
\label{eq:yBdefinition}
y_{\mathrm{B}} = \ln{(x_i/x_j)} = (y_1+y_2)/2 , 
\end{equation*}

\noindent where $y_{\mathrm{B}}$ is the rapidity of the boosted center-of-mass frame, $x_i$ and $x_j$ are the fractions of the proton momentum (Bjorken $x$) carried by each incident parton, and $y_1$ and $y_2$ are the rapidities of the outgoing partons in the detector frame.  The measured shapes of the observed $dN/d\chi$ distributions differ from the parton-level distributions because the observed ones convolve the parton-level distributions with non-uniform parton momentum distributions in $x_i$ and $x_j$, and also contain some admixture of non-$t$-channel processes. 
Restricting the range of the two-parton invariant mass and placing an upper bound on $y_{\mathrm{B}}$ 
reduces these differences.

The $dN/d\chi$ (angular) distributions of events with $|\yStar| < 1.7$
and $|y_{\mathrm{B}}| < 1.1$ are analyzed for contributions from BSM signals. 
The data with $\mjj< 2.5$~\TeV\xspace are discarded to remove trigger inefficiencies which otherwise arise due to the loosened $\yStar$ selection compared to the resonance analysis. The dataset is then analyzed by fitting to it a \textsc{Pythia} MC sample acting as an SM template as explained below. This sample is simulated as described in Section~\ref{sec:simulation}, including the aforementioned corrections.
Figure \ref{fig:two} shows the angular distributions of the data in different \mjj ranges starting from 3.4 \TeV, the SM prediction for the shape of the angular distributions after it is fit to data, and examples of the signals described in Section~\ref{sec:signals}. 
In the statistical analysis, MC simulation is normalized to data; in Figure~\ref{fig:two} both the MC simulation and the data are normalized to unit integral in each \mjj range for clarity of display.

Theoretical uncertainties in simulations of the angular distributions from QCD processes are estimated as described in Ref.~\cite{EXOT-2015-02}.\footnote{Uncertainties in electroweak corrections are not yet available and so are not included.}
The effect of varying the choice of PDF sets on the multijet prediction is estimated using \textsc{NLOJET}++ with three different PDF sets: CT10~\cite{Lai:2010vv}, MSTW2008~\cite{Martin:2009iq} and NNPDF2.3~\cite{Ball:2012cx}.
As the choice of PDF mainly affects the total cross-section rather than the shape of the $\chi$ distributions, these uncertainties are negligible ($<1\%$) in this analysis.
The uncertainty due to the choice of renormalization and factorization scales is estimated using \textsc{NLOJET}++ by varying each one independently up and down by a factor of two. 
The resulting uncertainties, taken as the variations in the normalized $\chi$ distributions, depend on both \mjj and $\chi$ and rise to 
12\%~(8\%) for the renormalization~(factorization) scale, at the smallest $\chi$ values and high \mjj values.
The statistical uncertainty in the simulated NLO corrections is less than 1\%.
The dominant experimental uncertainty in the predictions of the $\chi$ distributions is the jet energy scale uncertainty, with an impact of at most 15\% at high \mjj values, for the raw distribution before the fit is performed.
The uncertainty in the jet energy resolution has negligible impact.
The theoretical uncertainties and the total uncertainties are displayed as shaded bands around the prediction in Figure~\ref{fig:two}, where theoretical uncertainties can be seen to dominate.

\begin{figure}[b!]
 \centering 
 \includegraphics[width=\columnwidth]{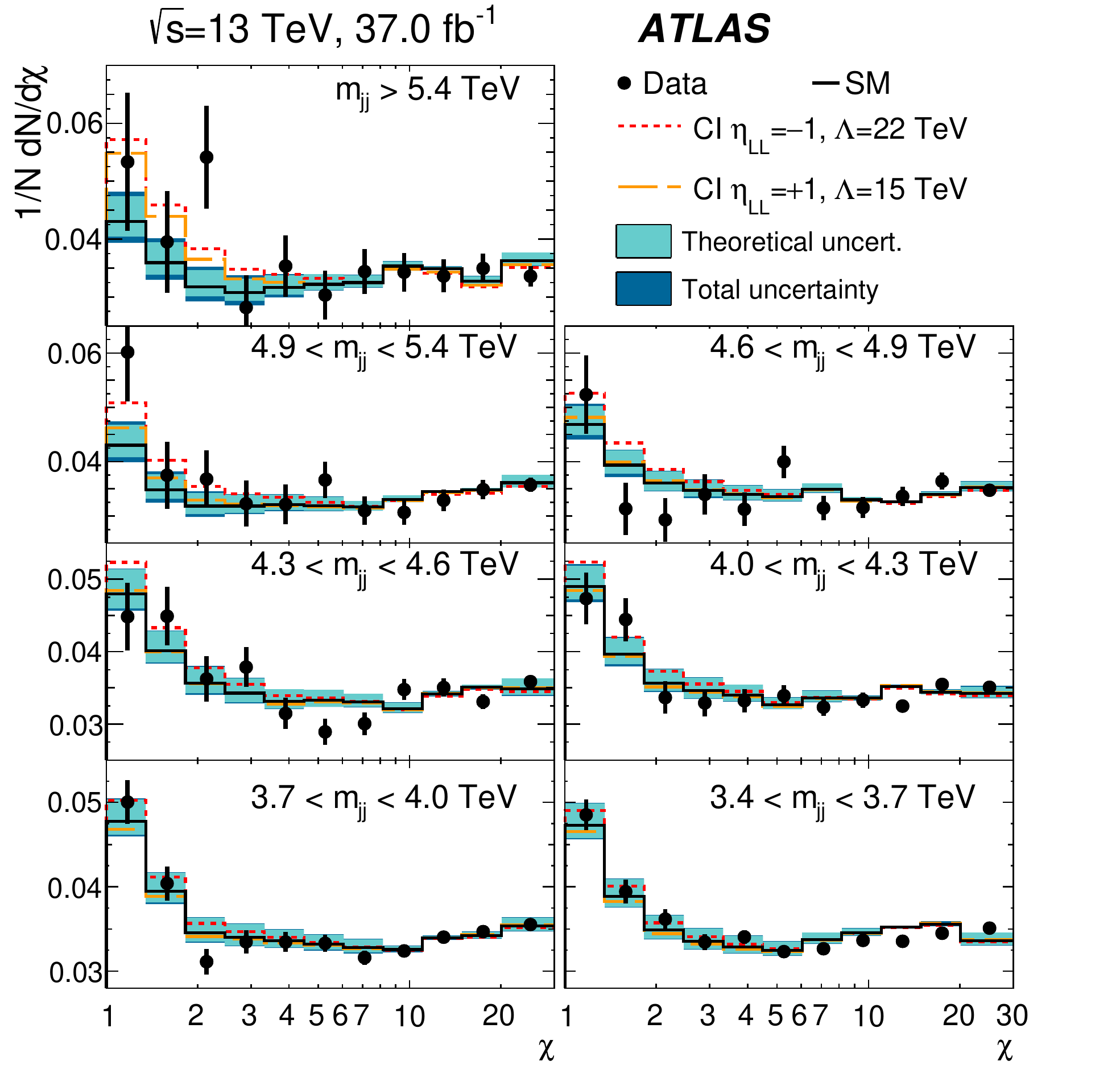}
  \caption{Reconstructed distributions of the dijet angular variable $\chi$ in
    different regions of the dijet invariant mass \mjj for events with
    $|\yStar|<1.7$, $|y_{\mathrm{B}}| < 1.1$, and $\pT>440$~$(60)$~$\GeV\xspace$
    for the leading~(subleading) jet. The data (points),
    \textsc{Pythia} predictions with NLO and electroweak corrections applied (solid lines), and examples of
    the contact interaction (CI)
    signals discussed in the text (dashed lines) are shown.
    The theoretical uncertainties and the total theoretical and experimental uncertainties in the predictions are displayed as
    shaded bands around the SM prediction.
    The SM background prediction and corresponding systematic uncertainty bands are extracted from the best-fit to the data. 
    Data and predictions are normalized to unity in each \mjj\ bin.}
  \label{fig:two}
\end{figure}

The compatibility of the $\chi$ distribution in data with the SM prediction and with the BSM signals discussed in Section~\ref{sec:signals} is tested using a combined fit in seven coarse \mjj\ bins covering $\mjj > 3.4$~TeV as shown in Figure~\ref{fig:two}.
The range $\mjj < 3.4$~\TeV\ provides no sensitivity to the studied benchmark models in ranges which are not yet excluded. A profile likelihood fit is performed, using as templates the $dN/d\chi$ distributions in each \mjj bin for data and QCD MC events. The MC simulation is normalized to the data separately in each \mjj bin, making this a shape-only comparison.
All systematic uncertainties are treated as correlated in \mjj; where this assumption is less secure, such as for the choice of MC event generator tune, other correlation models are tested and the differences are found to be inconsequential. The fit to the data is strongly constrained by the lowest \mjj bins, which have good statistical precision as well as negligible contributions from possible BSM signals, providing constraints of between 20\% and 40\% on the uncertainties in the higher $m_{jj}$ bins.
The~$CL_{b}$, or confidence level for the background-only hypothesis, comparing data to SM predictions is \CLsPValue. Thus no significant deviation of the data from the background-only hypothesis is observed. Limits on the production of BSM signals are set using the $CL_{s}$ method~\cite{Read:2002hq,Junk:1999kv}, which takes the $CL_{b}$ value into account and thereby avoids setting overly strong limits in light of the rather low observed $p$-value.

\section{Benchmark signals}
\label{sec:signals}

The data are used to constrain several of the many BSM models that predict dijet excesses. Excited quarks, quantum black holes, and $W^\prime$, \Wstar, and $Z^\prime$ bosons would produce peaks in the \mjj\xspace distribution. Contact interactions would introduce smooth changes in the high-mass tail of the $\mjj$ distribution that could be detected in the analysis of the $\chi$ distributions. 
The signal models are simulated using the parton-level event generators indicated below, in an identical manner to QCD processes, using the same PDFs and parameters for non-perturbative effects, except where noted otherwise.  The renormalization and factorization scales are set to the average \pt\ of the two leading jets.  The efficiency for all signal models is close to unity, henceforth acceptance times efficiency is referred to as acceptance.
For all models, acceptance is computed from all events which pass the analysis selection, including distribution tails caused by the sharp rise of PDFs at low Bjorken $x$.

If extra spatial dimensions exist, the fundamental scale of gravity could be lowered to a few \TeV\xspace and the LHC could produce quantum black holes at or above this scale~\cite{Giddings:2001bu,Dimopoulos:2001hw,RandallMeade,Feng:2004,Calmet:2008dg,Gingrich:2009hj,BoelaertAkesson:2010}.
High-multiplicity final states from thermalizing black holes are explored at $\sqrt{s} =13$~TeV by ATLAS in Refs.~\cite{EXOT-2015-09,EXOT-2015-14} and by CMS in Ref.~\cite{CMS-EXO-12-009}.
This analysis explores QBH that would be produced at or above the fundamental scale of gravity $M_{\rm D}$ and decay into a few particles rather than the high-multiplicity final states characteristic of thermalizing black holes~\cite{Gingrich:2009hj,Calmet:2008dg,Dai:2007ki,Frost:2009cf}. These would appear in the \mjj distribution as an excess localized near the threshold mass for quantum black hole production, \mthreshold. 
Here, production and decay to two jets is simulated using the \textsc{BlackMax} event generator~\cite{Dai:2007ki}
 assuming an Arkani-Hamed\discretionary{--}{}{--}Dimopoulous\discretionary{--}{}{--}Dvali (ADD) scenario~\cite{ArkaniHamed:1998rs,Antoniadis:1998ig} with $M_{\rm D}=\mthreshold$ and a number of extra dimensions $n = 6$, as in Ref.~\cite{EXOT-2013-11}. 
In this model, the branching ratio to dijets is greater than 96\%.
The PDFs used are CTEQ6L1~\cite{Pumplin:2002vw}. The QBH signals peak slightly above their threshold values and have negligible low-mass tails. 
The reconstructed signal peaks have width-to-mass ratios of approximately 10\%.
The acceptance of the resonance search selection for quantum black holes is approximately \blackMaxAcceptance\ across all studied masses. 

Excited quarks are predicted in models of compositeness and are a typical benchmark for quark--gluon resonances used in many past dijet searches~\cite{CDF:2009DijetSearch,EXOT-2010-01,CMS-EXO-10-010,CMS-EXO-15-001,EXOT-2015-02}.
The $q^*$ model is simulated with \textsc{Pythia~8.186}, assuming spin-1/2 excited quarks with coupling constants the same as for SM quarks; no interference with the SM is simulated.
Only the decay of the excited quark to a gluon and an up- or down-type quark is simulated; this corresponds to a branching ratio of 85\%.
Before parton shower effects are taken into account,
the intrinsic width of the $q^*$ signals is comparable to the detector resolution. After showering, a radiative tail is present that increases in strength for higher \qstar\ masses, an effect augmented by the impact of PDFs decreasing towards higher masses.
The resonance search selection acceptance for a $q^*$ with a mass of 4~\TeV\xspace is \qstarAcceptance.

Additional spin-1 $W^\prime$ and $Z^\prime$ bosons often arise in the symmetry breaking of extended gauge theories.
A $W^\prime$ model with axial-vector SM couplings 
and a corresponding branching ratio to quarks of 75\% is considered~\cite{Langacker:1984dc}.
Events are simulated with \textsc{Pythia~8.205} and decays 
are restricted to quark--antiquark pairs with all three quark-flavor doublets included.
A leptophobic $Z^\prime$ model is also simulated, with matrix elements calculated in \textsc{MadGraph5\_aMC@NLO} v2.2.3~\cite{Alwall:2014hca} and parton showering performed in \textsc{Pythia~8.210}.
The $Z^\prime$ model assumes axial-vector couplings to all SM quarks and to a Dirac fermion dark matter candidate. 
Final states with top quarks are not simulated. 
The model considered follows a scenario~\cite{Chala:2015ama} where the $Z^\prime$ branching ratio to dark matter is negligible, 
hence the dijet production rate and resonance width depend only on the coupling to quarks, $g_q$, and the mass of the resonance $m_{Z^\prime}$.
Before parton shower effects are considered,
the intrinsic width of the $Z^\prime$ signal ranges from 0.05\% of the mass of a 1.5~\TeV\xspace $Z^\prime$ with $g_q=0.1$ to
10\% of the mass of a 3.5~\TeV\xspace $Z^\prime$ with $g_q=0.5$. The $W^\prime$ signal has an intrinsic width similar to a $Z^\prime$ of coupling $g_q = 0.3$ at every mass point considered.
For coupling values of $g_q = 0.6$ and above, the intrinsic width of the \Zprime\ for the mass range of interest increases to 15\% and beyond, resulting in a very wide peak and in a loss of sensitivity in the resonance search, which is therefore limited to $g_q\leq 0.5$.
No interference with the SM is simulated for either the $W^\prime$ or the $Z^\prime$ model.  
The resonance search selection acceptance for a mass of 3~\TeV\xspace is \wprimeAcceptance for the $W^\prime$ model and \zprimepttwpAcceptance for the $Z^\prime$ model with $g_q=0.2$. Because of the large radiative tails of the \Wprime\ signals, the acceptance for this model increases to a maximum at approximately 2.5~\TeV\ and decreases to values smaller than $20\%$ for masses above 6.0~\TeV.

An excited \Wstar\ boson is generated through
a simplified model \cite{Chizhov:2010ry} in the CalcHEP 3.6 event generator~\cite{Belyaev:2012qa},
in combination with the NNPDF2.3 NLO PDF set and
\textsc{Pythia~8.210} for the simulation of non-perturbative effects.
The mixing angle in this model ($\phi_{X}$)
is set to zero, producing leptophobic decays of the \Wstar\ 
that are limited to all SM quarks. The angular distribution of the \Wstar\ differs from
that of the other signals under study, peaking at \yStar\ values above 1. 
Therefore, this benchmark model is constrained using the alternative signal region with
$|\ystar| < 1.2$. The acceptance for the 
leptophobic \Wstar\ signal with this selection increases from \WstarAcceptance around 2 TeV to nearly 60\% for the highest masses examined.

Results are also provided as limits on the cross-section times acceptance times branching ratio to two jets,
$\sigma\times A\times\text{BR}$, of 
a hypothetical signal modeled as a Gaussian peak in the 
particle-level \mjj distribution.
When limits are set on Gaussian signal models that can contribute to the reconstructed \mjj~spectrum~(e.g. as in Ref.~\cite{EXOT-2013-11}),
the description of the corresponding distribution folds together the actual physical signal and detector effects~(acceptance and resolution).
Here a model is defined at particle level, within a fiducial region.
This model is then folded with the effects of the detector response,
described through an MC-based transfer matrix that relates the particle level and reconstructed observables.
The transfer matrix accounts for bin-to-bin migrations due to resolution effects, as well as for the fractions of events passing the selection only at particle or reconstruction level.
In order to avoid large simulation-based extrapolations, the fiducial selection at particle level matches the one applied at reconstruction level.
Limits on a given signal model can be interpreted from the phenomenological point of view at particle level,
without need for further information about the detector response.

For sufficiently narrow resonances, these results may be used to set limits in BSM models beyond those considered explicitly in this \publicationType. 
The predicted signals should be compared at particle level, after applying the resonance selection, with the limit that corresponds most closely to the width of the Gaussian contribution predicted by the model. 
Since a Gaussian signal shape is assumed in determining the limits, any long tails in the \mjj distribution should not be included in the model under study. 
A procedure similar to the one detailed in Appendix A.1 of Ref.~\cite{EXOT-2013-11} can be followed, after applying the non-perturbative corrections and performing the fiducial selection at particle level, without applying any further detector smearing as it is already accounted for in the folding procedure.

The folding procedure applied for the various signal samples discussed above, using transfer matrices based on either the same or different samples,
yields reconstructed distributions compatible with the ones from full simulation.
The limits on narrow signals at particle level, folded with the detector effects,
are similar to the ones obtained for a Gaussian signal at reconstruction level having a width
equal to the one expected from detector resolution.\footnote{ Differences of about $4\%$ between these limits are seen, due to non-Gaussian tails of the resolution which are taken into account by the folding matrix, but are not accounted for in the case of the Gaussian signal at reconstruction level. }
For resonance widths comparable to the resolution, differences up to about $20\%$ are observed between the results of the two limit-setting approaches.
The folding method yields results at particle level, accounting also for the mass dependence of the resolution within the range of the resonance, hence its relevance for providing results that are easy to interpret.
For large signal widths, the effect of the detector resolution on the global width is smaller and the difference between the results of the two limit-setting approaches is reduced.

For all signals described above, the following systematic uncertainties are included in the limit setting: jet energy scale, acceptance uncertainties associated to the choice of PDF, and luminosity.
The jet energy uncertainty ranges from 1.5\% at the lowest masses to 3\% for masses above 4.5 TeV.
On average, the PDF uncertainty affects the angular distributions by 1\%. The uncertainty in the combined 2015+2016 integrated luminosity is 3.2\%. It is derived, following a methodology similar to that detailed in Ref.~\cite{DAPR-2013-01}, from a preliminary calibration of the luminosity scale using $x$--$y$ beam-separation scans performed in August 2015 and May 2016. 

The dijet angular distributions can also be modified by new mediating particles with a mass much higher than that which can be probed directly. A four-fermion effective field theory (contact interaction) characterized by a single energy scale $\Lambda$ can be used to describe these effects:
\begin{align} 
\mathcal{L}_{qq} =  \frac{2\pi}{\Lambda^2} [&~\eta_{\mathrm{LL}}(\bar{q}_{\rm L}\gamma^\mu q_{\rm L})(\bar{q}_{\rm L}\gamma_\mu q_{\rm L}) \nonumber \\ 
&+ \eta_{\mathrm{RR}}(\bar{q}_{\rm R}\gamma^\mu q_{\rm R})(\bar{q}_{\rm R}\gamma_\mu q_{\rm R}) \nonumber \\
&+ 2\eta_{\mathrm{RL}}(\bar{q}_{\rm R}\gamma^\mu  q_{\rm R})(\bar{q}_{\rm L}\gamma_\mu q_{\rm L}) ]~,
\end{align}
\noindent
where the quark fields have left-handed (L) and right-handed (R) chiral projections and the coefficients $\eta_{\mathrm{LL}}$, $\eta_{\mathrm{RR}}$, and $\eta_{\mathrm{RL}}$ activate various interactions. 
Contact interactions with a non-zero left-chiral color-singlet coupling ($\eta_{\mathrm{LL}}= \pm 1$, $\eta_{\mathrm{RL}}=\eta_{\mathrm{RR}}= 0$) are simulated using \textsc{Pythia~8.186}. 
This type of coupling is chosen because its angular distributions are representative of those of other BSM models~(e.g. $Z^\prime $ and others studied here by the resonance search).
Interference of the signal model with the SM process $q\bar{q}~\rightarrow~q\bar{q}$ is included.
Events are simulated for both constructive and destructive interference with $\Lambda = 7$~\TeV\xspace. 
From this sample, the angular distributions for other values of $\Lambda$ are obtained using the fact that the interference term is proportional to $1/\Lambda^2$ and the pure contact-interaction cross-section is proportional to $1/\Lambda^4$.
The \textsc{Pythia} signal prediction is reweighted to the NLO cross-sections provided by CIJET~\cite{CIJET}.
Uncertainties in the prediction of the angular distributions for contact interaction signals are obtained in the same manner as for QCD processes, including JES and PDF uncertainties (as discussed in Section~\ref{sec:angular}).

\section{Results}
\label{sec:results}

Starting from the \mjj distribution obtained with the resonance search selection, a Bayesian method~\cite{EXOT-2010-07} is applied to the data and simulation of signals at a series of discrete masses to set 95\% credibility-level (CL) upper limits on the cross-section times acceptance for the signals described above.
The method uses a constant prior for the signal cross-section and Gaussian priors for nuisance parameters corresponding to systematic uncertainties in the signal and background distributions.
The expected limits are calculated using pseudo-experiments generated from the maximum-likelihood values of the background uncertainties in the sliding-window background model and accounting for the full set of systematic uncertainties in both the signal and background models.
The limit is interpolated logarithmically between the discrete masses to create continuous exclusion curves. 
No uncertainty in the theoretical cross-section for the signals is assessed.
The various selection criteria for the different signal regions are summarized in Table~\ref{tab:CutSummary}.
The mass limits for each of the models are shown in Figures~\ref{fig:limitsModels} and \ref{fig:zprimeTwoD} and Table~\ref{tab:Limits}.

Figure \ref{fig:four} shows limits on the Gaussian contributions to the particle-level \mjj distribution obtained for a mean mass $m_G$ and five different widths, from a narrow width to a width of 15\% of $m_G$.
The expected limit and the corresponding $\pm 1 \sigma$ and $\pm 2\sigma$ bands are also indicated for a narrow-width resonance.
Limits are set only when $m_G$ is within $1.1\text{--}6.5$~\TeV~and separated by at least the width of the Gaussian resonance from the beginning of this range.
Resonances with effective cross-sections exceeding values ranging \gaussianRangeStatement are excluded.
As the width increases, the expected signal contribution is distributed across more bins.
Therefore, wider signals are less affected by statistical fluctuations of the data in a single bin than narrower signals.

Starting from the $\chi$ distributions obtained with the angular selection, the $CL_{s}$ method is used to set limits on
potential contributions from contact interactions, using the background predicted by the SM simulation as the null hypothesis.
The asymptotic approximation ~\cite{Cowan:2010js} of a profile likelihood ratio is used to set 95\% CL limits.
For each value of $\Lambda$ and each $\eta_{\text{LL}}$ tested, a combined fit is performed on the seven $m_{jj}$ regions of Figure~\ref{fig:two}, using the procedure described in Section \ref{sec:angular}.
The maximum-likelihood values of the nuisance parameters do not differ significantly from the expectations.
The bounds on contact interactions thus obtained are shown in Figure~\ref{fig:six} and in Table~\ref{tab:Limits}. In the case of destructive interference, the expected event yield including the signal may be lower than that for the background-alone prediction. The kinematic regions where this occurs depend on both $\Lambda$ and $\mjj$.
An observed excess in the data then produces a weaker limit below a given $\Lambda$ value, and a stronger one above that $\Lambda$ value, in combination with information from the $\mjj$ spectrum in the fit.

The same approach is used to set limits on the resonant benchmark signals described in Section \ref{sec:signals}, 
as a consistency check of the resonance search. The angular analysis has a $\sim$10\% lower
sensitivity in terms of resonance mass exclusion with respect to the search described in Section \ref{sec:resonant}.

\clearpage

\begin{table}[!h]
	\centering
	\caption{Summary of the analysis selection criteria for the three considered signal regions.}
	\begin{tabular}{l c c c c c }
		\toprule
		& $\pt^{\rm{leading}}$ & $\pt^{\rm{subleading}}$ &  $|\ystar|$  & $|y_{\text{B}}|$ & $\mjj$  \\
		\midrule
		Resonance & > 0.44 \TeV & > 0.06 \TeV & < 0.6  & - & > 1.1 \TeV   \\
		
		\Wstar & > 0.44 \TeV & > 0.06 \TeV & < 1.2  & - & > 1.7 \TeV   \\
		
		Angular & > 0.44 \TeV & > 0.06 \TeV & < 1.7  & < 1.1 & > 2.5 \TeV   \\
		\bottomrule
	\end{tabular}
	\label{tab:CutSummary}
\end{table}

\begin{table}[!h]
	\centering
	\caption{The 95\% CL lower limits on the masses of ADD quantum black holes (\textsc{\BlackMax} event generator), \Wprime\ and \Wstar\ bosons, excited quarks, and $Z^\prime$ bosons for selected coupling values from the resonance search, as well as on the scale of contact interactions for constructive ($\eta_{\text{LL}}=-1$) and destructive ($\eta_{\text{LL}}=+1$) interference from the angular analysis. Where an additional range is listed, masses within the range are also excluded. Full limits on the $Z^\prime$ model are provided in Figure~\ref{fig:zprimeTwoD}.}
	\begin{tabular}{ l c c}
		\toprule
		\multicolumn{1}{c}{Model} & \multicolumn{2}{c}{95\% CL exclusion limit} \\
		\cmidrule(r){2-3}
		&  Observed & Expected \\
		\midrule
		Quantum black hole &  \blackMaxLimitMjj  & \blackMaxLimitMjjExpected \\
		\midrule
		\Wprime  & \wprimeLimitMjj &   \wprimeLimitMjjExpected \\
		\midrule
		\multirow{2}{*}{\Wstar}  & \WstarLimitMjjOne  &    \multirow{2}{*}{~\WstarLimitMjjExpected}  \\
		& \WstarLimitMjjTwo & \\		
		\midrule
		Excited quark & \qstarLimitMjj  &   \qstarLimitMjjExpected  \\
		\midrule
		\Zprime ($g_q= 0.1$)  & \zprimeLowLimitMjj & \zprimeLowLimitMjjExpected \\
		\midrule
		\Zprime ($g_q= 0.2$)  & \zprimeHighLimitMjj & \zprimeHighLimitMjjExpected \\
		\midrule                                               
		Contact interaction ($\eta_{\text{LL}}= -1$) & \LambdaLimitMinusChi\phantom{.} & \LambdaLimitMinusChiExpected\phantom{.} \\
		\midrule
		\multirow{2}{*}{Contact interaction ($\eta_{\text{LL}}= +1$)}  & \LambdaLimitPlusChi\phantom{.} & \multirow{2}{*}{\LambdaLimitPlusChiExpected\phantom{.}} \\
		& \LambdaObsDestrMin\ -- \LambdaObsDestrMax & \\
		\bottomrule
	\end{tabular}
	\label{tab:Limits}
\end{table}

\clearpage

\begin{figure}[h!]
	\centering
	\subfigure[\qstar]{\includegraphics[width=0.45\columnwidth]{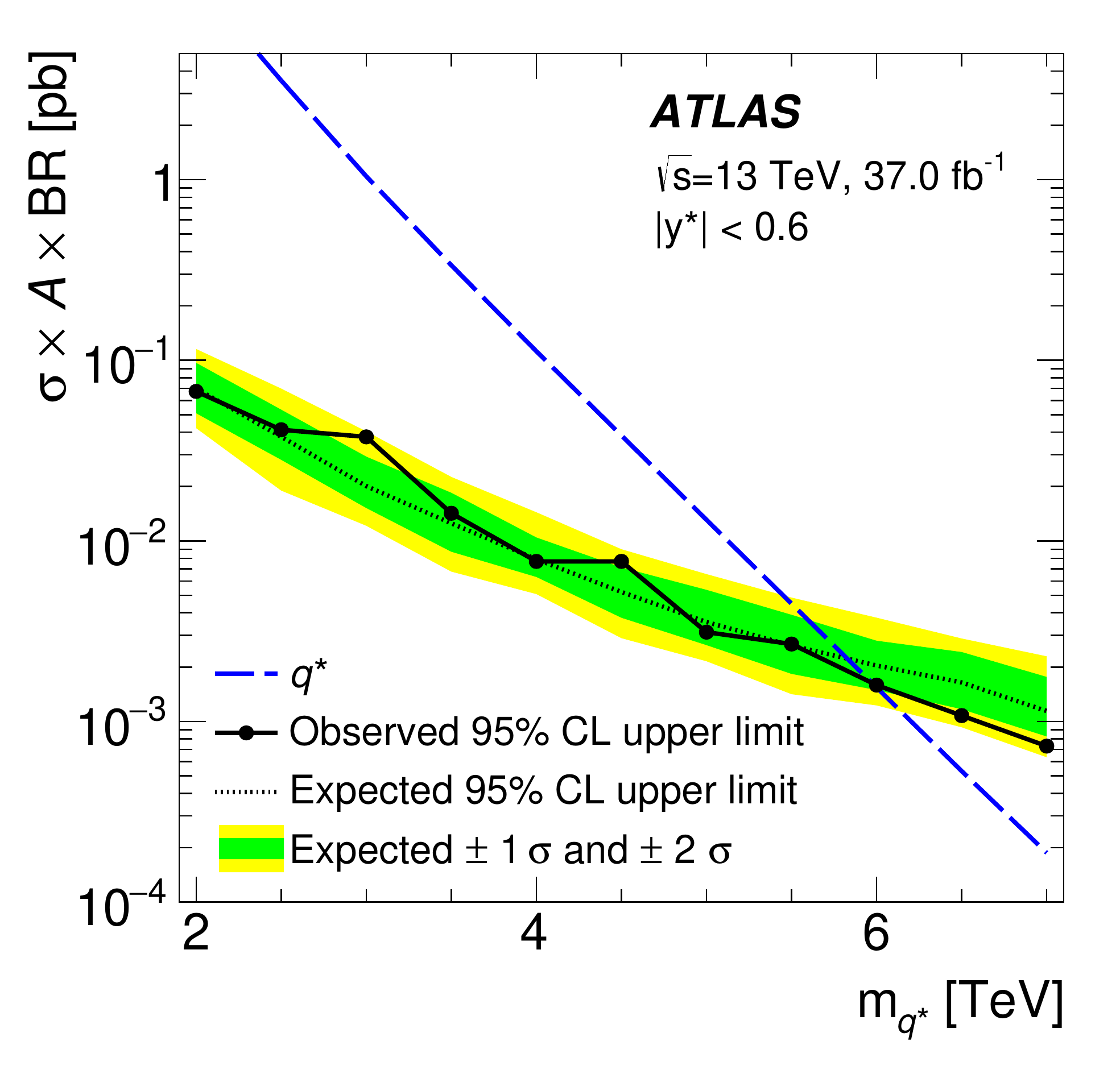}\label{subfig:qstarlimit}}
	\subfigure[QBH]{\includegraphics[width=0.45\columnwidth]{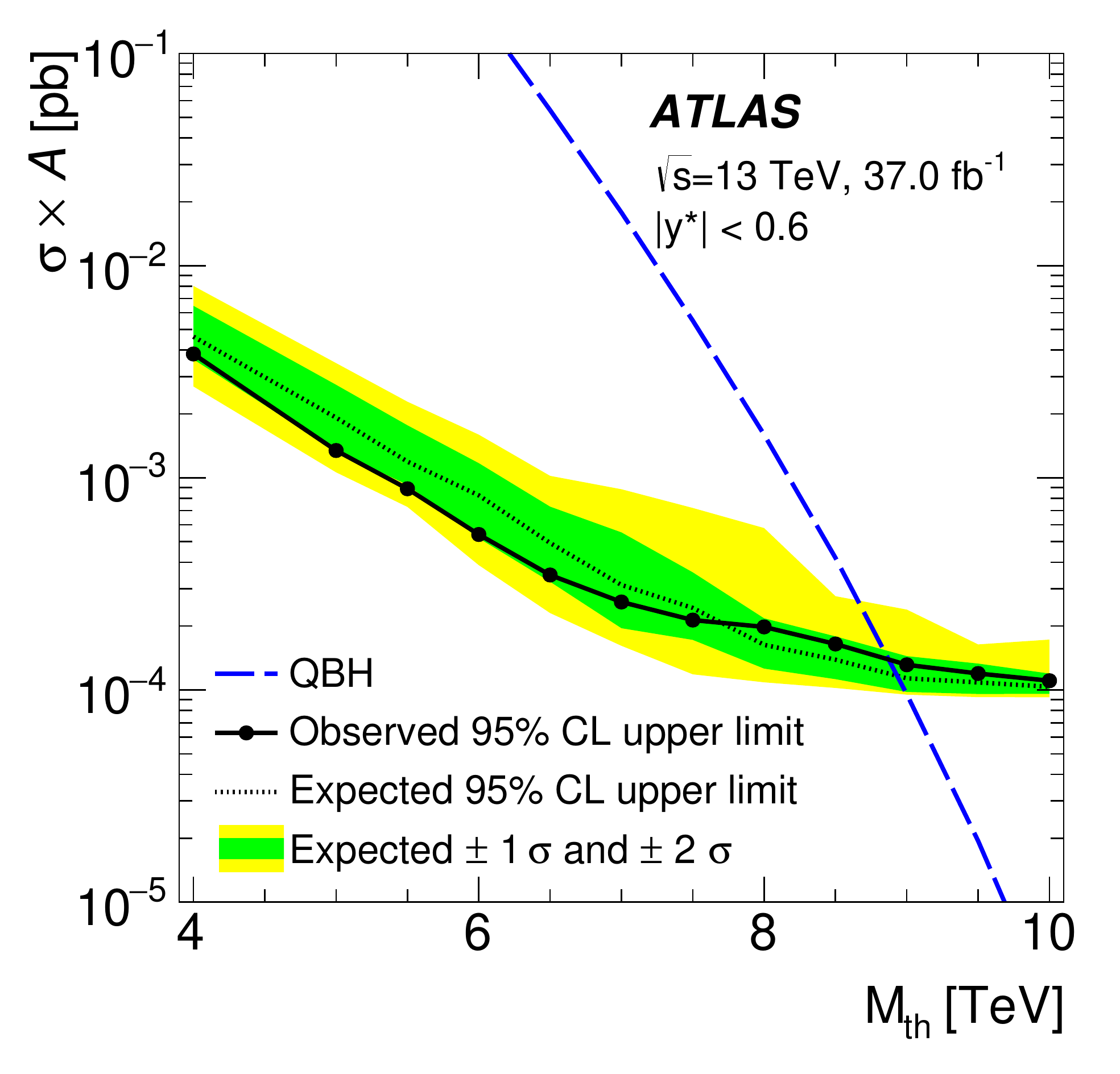}\label{subfig:blackmax}}
	\subfigure[\Wprime]{\includegraphics[width=0.45\columnwidth]{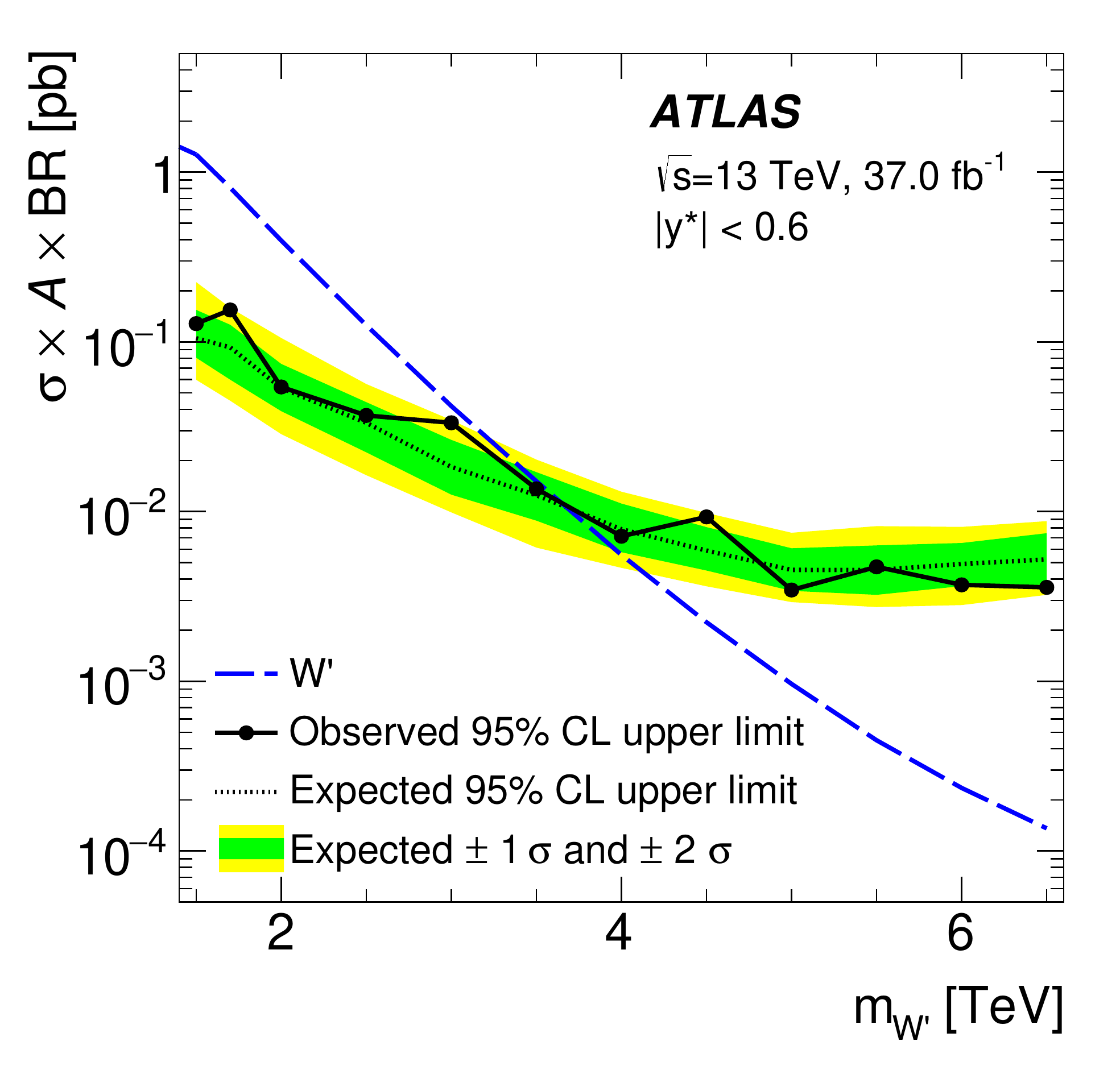}\label{subfig:wprime}}
	\subfigure[\Wstar]{\includegraphics[width=0.45\columnwidth]{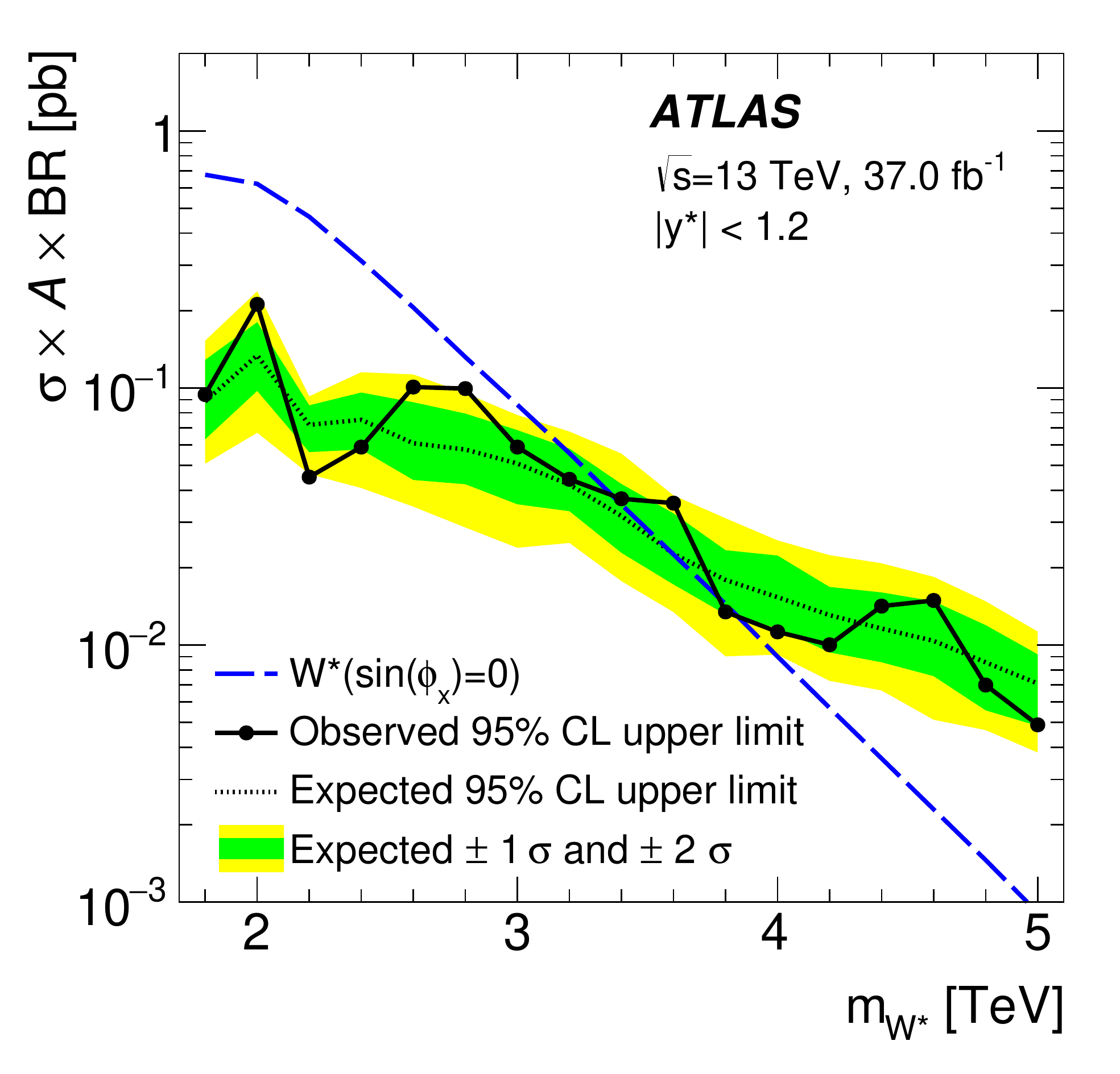}\label{subfig:wstar}}
	\caption{The 95\% CL upper limits obtained from the dijet invariant mass (\mjj) distribution on cross-section times acceptance times branching ratio to two jets, $\sigma\times A\times\text{BR}$, for the models described in the text. Clockwise from top left: \qstar,
		quantum black holes with $n=6$ generated with \BlackMax, 
		\Wprime, and \Wstar\, where the first three use the nominal selection and the last uses the widened $|y^{\ast}| < 1.2$ selection.  
		The numerical values of the observed and expected limits are summarized in Table~\ref{tab:Limits}.
	}
	\label{fig:limitsModels}
\end{figure}

\begin{figure}[h!]
	\centering
	\includegraphics[width=1.00\columnwidth]{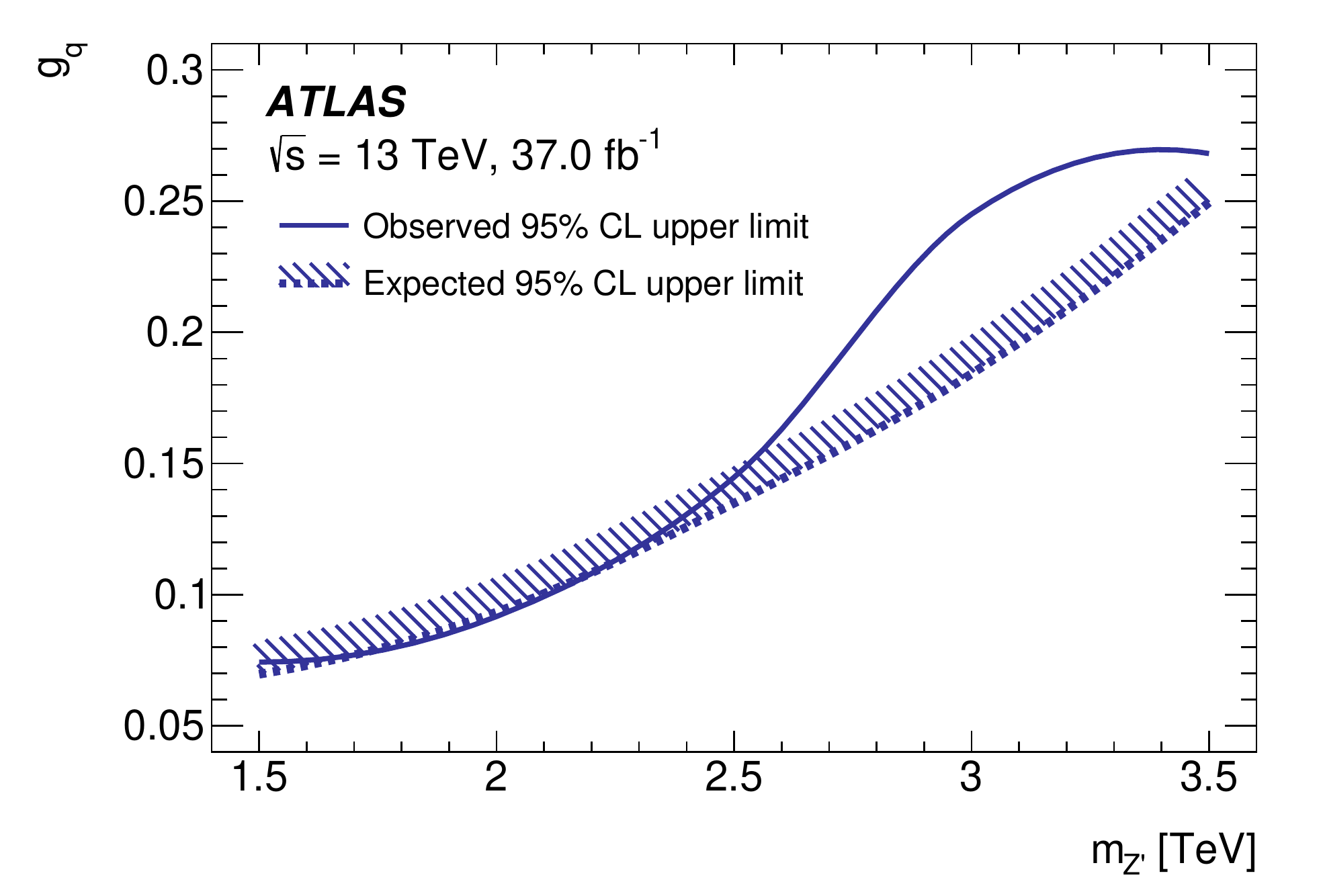}
	\caption{The 95\% CL exclusion limits for the $Z^{\prime}$ model described in the text,
		as a function of the coupling to quarks, $g_q$, and the mass, $m_{Z^\prime}$, 
		obtained from the dijet invariant mass \mjj distribution. For a given mass, the cross-sections rise with $g_q$, and thus the upper left unfilled area is excluded, as indicated by the direction of the hatched band.
		The exclusion applies up to $g_q=0.5$, in the sensitivity range of the method as explained in the text.
		Points were simulated with 0.5 TeV spacing in mass and spacing as fine as 0.05 in $g_q$.
		A smooth curve is drawn between points by interpolating in $g_q^2$ followed by an interpolation in $m_{Z^\prime}$.
	}
	\label{fig:zprimeTwoD}
\end{figure}

\begin{figure}[h!]
	\centering
	\includegraphics[width=1.00\columnwidth]{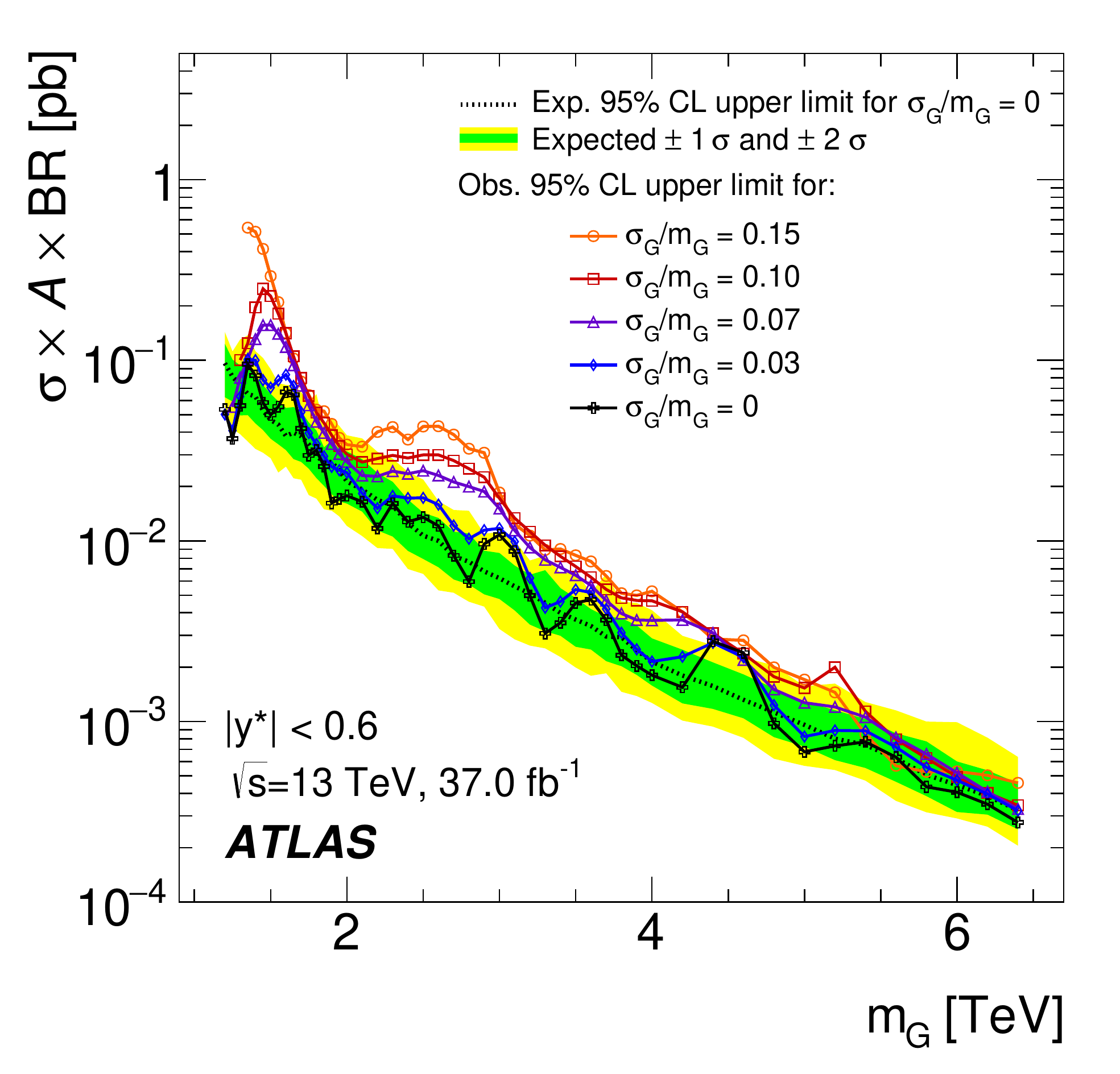}
	\caption{The 95\% CL upper limits obtained from the dijet invariant mass \mjj distribution on cross-section times acceptance times branching ratio to two jets, $\sigma\times A\times\text{BR}$, for a hypothetical signal with a cross-section $\sigma_G$ that produces a Gaussian contribution to the particle-level \mjj distribution, as a function of the mean of the Gaussian mass distribution $m_G$. Observed limits are obtained for five different widths, from a narrow width to 15\% of $m_G$. The expected limit and the corresponding $\pm 1 \sigma$ and $\pm 2\sigma$ bands are also indicated for a narrow-width resonance.
	}
	\label{fig:four}
\end{figure}

\begin{figure}[!htb]
	\centering
	\subfigure{\includegraphics[width=1.00\columnwidth]{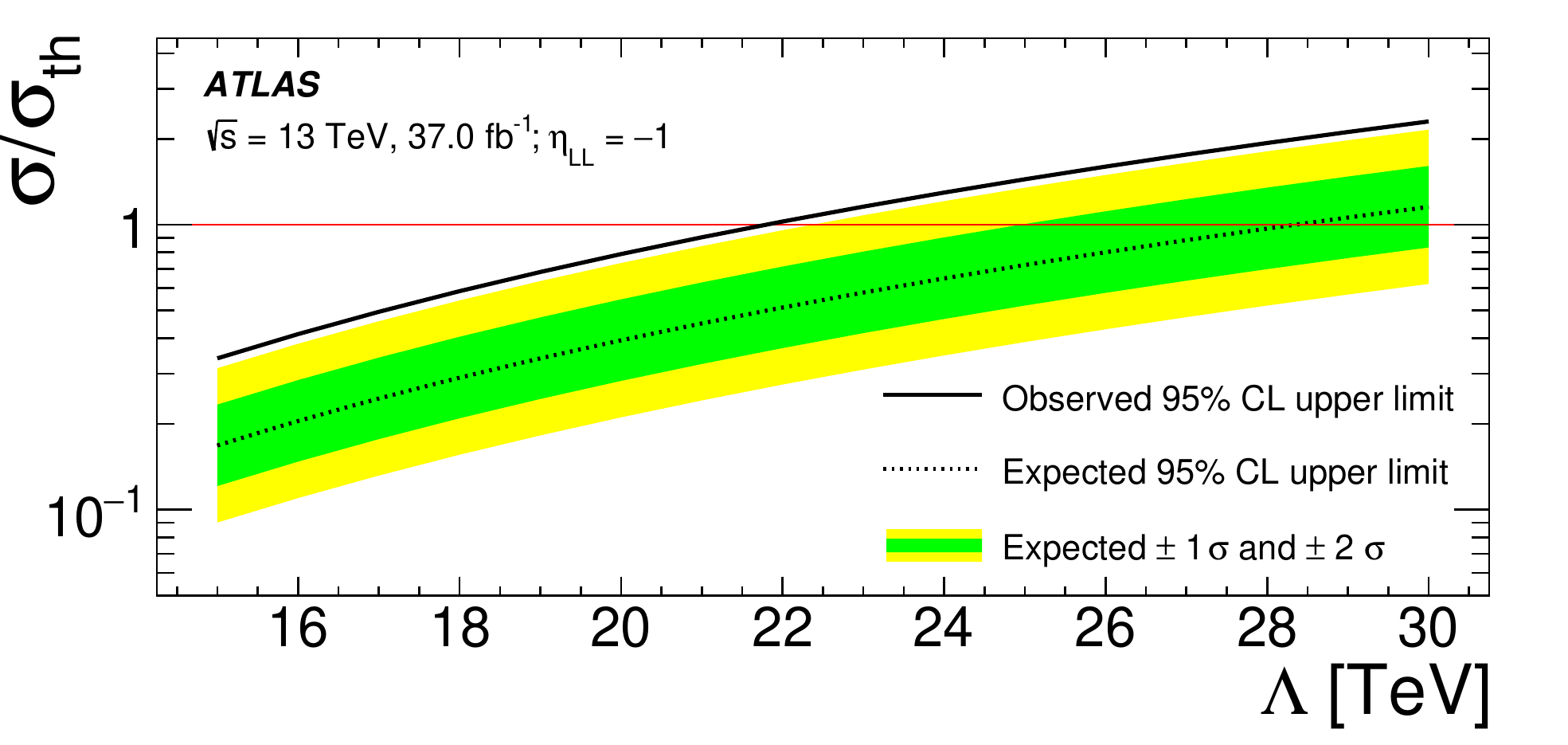}\label{subfig:constructive}}
	\subfigure{\includegraphics[width=1.00\columnwidth]{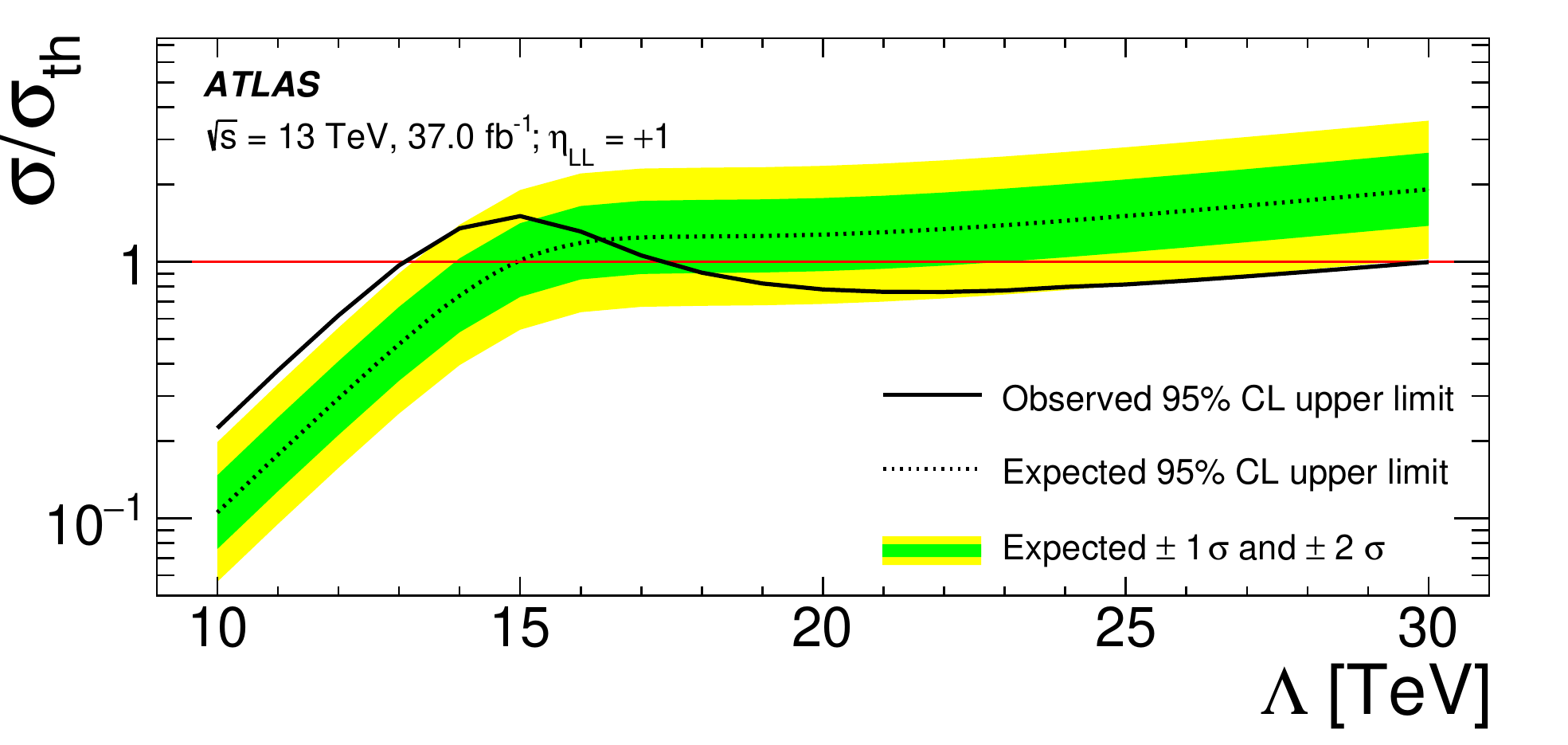}
		\label{subfig:destructive}}
	\caption{Ratio $\sigma/\sigma_{\text{th}}$ of the observed and expected 95\% CL upper limits on the cross-section in the contact interaction model to the predicted cross-section as a function of the compositeness scale $\Lambda$, for constructive (top) and destructive (bottom) interference with QCD processes. The $\Lambda$ regions for which the observed and expected 95\% CL lines are below the line at 1.0 represent the observed and expected exclusion regions, respectively.
		The numerical values of the observed and expected limits are summarized in Table~\ref{tab:Limits}.}
	\label{fig:six}
\end{figure}

 \FloatBarrier

\section{Conclusion}
\label{sec:conclusion}

A search for new phenomena beyond the Standard Model has been performed using dijet events in \integLumi\ of proton--proton collisions with a center-of-mass energy of $\sqrt{s}=13$~TeV recorded by the ATLAS detector at the Large Hadron Collider.
The dijet invariant mass distribution exhibits no significant local excesses above a data-derived estimate of the smoothly falling distribution predicted by the Standard Model. 
The two resonant signal regions agree with the background-only hypothesis, with $p$-values of \bumpHunterPValue and \bumpHunterPValueWStar for the $|\yStar| < 0.6$ and $|\yStar| < 1.2$ selections respectively. 
The dijet angular distributions, based on the rapidity difference between the two leading jets, also agree with a MC simulation of the SM, with a $p$-value for the SM-only hypothesis of \CLsPValue.
With the resonance selection, the analysis excludes several types of signals at 95\% CL, as predicted by models of quantum black holes, excited quarks, and \Wprime, \Wstar\, and \Zprime\ bosons. 
It also sets 95\% CL upper limits on the cross-section for new processes that would produce a Gaussian contribution to the dijet mass distribution.
With the angular analysis, 95\% CL lower limits are set on the compositeness scale of contact interactions for scenarios with either constructive or destructive interference between the new interaction and QCD processes.
These results substantially extend the excluded ranges obtained using the 2015 dataset alone, with improvements ranging from 7\% for quantum black hole masses to 25\% for contact interaction scales to 40\% for \Wprime\ boson masses.

\section*{Acknowledgements}
\label{sec:acknowldge}

We thank CERN for the very successful operation of the LHC, as well as the
support staff from our institutions without whom ATLAS could not be
operated efficiently.

We acknowledge the support of ANPCyT, Argentina; YerPhI, Armenia; ARC, Australia; BMWFW and FWF, Austria; ANAS, Azerbaijan; SSTC, Belarus; CNPq and FAPESP, Brazil; NSERC, NRC and CFI, Canada; CERN; CONICYT, Chile; CAS, MOST and NSFC, China; COLCIENCIAS, Colombia; MSMT CR, MPO CR and VSC CR, Czech Republic; DNRF and DNSRC, Denmark; IN2P3-CNRS, CEA-DSM/IRFU, France; SRNSF, Georgia; BMBF, HGF, and MPG, Germany; GSRT, Greece; RGC, Hong Kong SAR, China; ISF, I-CORE and Benoziyo Center, Israel; INFN, Italy; MEXT and JSPS, Japan; CNRST, Morocco; NWO, Netherlands; RCN, Norway; MNiSW and NCN, Poland; FCT, Portugal; MNE/IFA, Romania; MES of Russia and NRC KI, Russian Federation; JINR; MESTD, Serbia; MSSR, Slovakia; ARRS and MIZ\v{S}, Slovenia; DST/NRF, South Africa; MINECO, Spain; SRC and Wallenberg Foundation, Sweden; SERI, SNSF and Cantons of Bern and Geneva, Switzerland; MOST, Taiwan; TAEK, Turkey; STFC, United Kingdom; DOE and NSF, United States of America. In addition, individual groups and members have received support from BCKDF, the Canada Council, CANARIE, CRC, Compute Canada, FQRNT, and the Ontario Innovation Trust, Canada; EPLANET, ERC, ERDF, FP7, Horizon 2020 and Marie Sk{\l}odowska-Curie Actions, European Union; Investissements d'Avenir Labex and Idex, ANR, R{\'e}gion Auvergne and Fondation Partager le Savoir, France; DFG and AvH Foundation, Germany; Herakleitos, Thales and Aristeia programmes co-financed by EU-ESF and the Greek NSRF; BSF, GIF and Minerva, Israel; BRF, Norway; CERCA Programme Generalitat de Catalunya, Generalitat Valenciana, Spain; the Royal Society and Leverhulme Trust, United Kingdom.

The crucial computing support from all WLCG partners is acknowledged gratefully, in particular from CERN, the ATLAS Tier-1 facilities at TRIUMF (Canada), NDGF (Denmark, Norway, Sweden), CC-IN2P3 (France), KIT/GridKA (Germany), INFN-CNAF (Italy), NL-T1 (Netherlands), PIC (Spain), ASGC (Taiwan), RAL (UK) and BNL (USA), the Tier-2 facilities worldwide and large non-WLCG resource providers. Major contributors of computing resources are listed in Ref.~\cite{ATL-GEN-PUB-2016-002}.


\clearpage
\printbibliography

\newpage 
\begin{flushleft}
{\Large The ATLAS Collaboration}

\bigskip

M.~Aaboud$^\textrm{\scriptsize 137d}$,
G.~Aad$^\textrm{\scriptsize 88}$,
B.~Abbott$^\textrm{\scriptsize 115}$,
J.~Abdallah$^\textrm{\scriptsize 8}$,
O.~Abdinov$^\textrm{\scriptsize 12}$$^{,*}$,
B.~Abeloos$^\textrm{\scriptsize 119}$,
S.H.~Abidi$^\textrm{\scriptsize 161}$,
O.S.~AbouZeid$^\textrm{\scriptsize 139}$,
N.L.~Abraham$^\textrm{\scriptsize 151}$,
H.~Abramowicz$^\textrm{\scriptsize 155}$
% ...
\bigskip
\\
% ...
$^{8}$ Department of Physics, The University of Texas at Arlington, Arlington TX, United States of America\\
% ...
$^{12}$ Institute of Physics, Azerbaijan Academy of Sciences, Baku, Azerbaijan\\
% ...
$^{88}$ CPPM, Aix-Marseille Universit{\'e} and CNRS/IN2P3, Marseille, France\\
% ...
$^{115}$ Homer L. Dodge Department of Physics and Astronomy, University of Oklahoma, Norman OK, United States of America\\
% ...
$^{119}$ LAL, Univ. Paris-Sud, CNRS/IN2P3, Universit{\'e} Paris-Saclay, Orsay, France\\
% ...
$^{137}$ $^{(a)}$ Facult{\'e} des Sciences Ain Chock, R{\'e}seau Universitaire de Physique des Hautes Energies - Universit{\'e} Hassan II, Casablanca; $^{(b)}$ Centre National de l'Energie des Sciences Techniques Nucleaires, Rabat; $^{(c)}$ Facult{\'e} des Sciences Semlalia, Universit{\'e} Cadi Ayyad, LPHEA-Marrakech; $^{(d)}$ Facult{\'e} des Sciences, Universit{\'e} Mohamed Premier and LPTPM, Oujda; $^{(e)}$ Facult{\'e} des sciences, Universit{\'e} Mohammed V, Rabat, Morocco\\
% ...
$^{139}$ Santa Cruz Institute for Particle Physics, University of California Santa Cruz, Santa Cruz CA, United States of America\\
% ...
$^{151}$ Department of Physics and Astronomy, University of Sussex, Brighton, United Kingdom\\
% ...
$^{155}$ Raymond and Beverly Sackler School of Physics and Astronomy, Tel Aviv University, Tel Aviv, Israel\\
% ...
$^{161}$ Department of Physics, University of Toronto, Toronto ON, Canada\\
% ...
$^{a}$ Also at Department of Physics, King's College London, London, United Kingdom\\
$^{b}$ Also at Institute of Physics, Azerbaijan Academy of Sciences, Baku, Azerbaijan\\
$^{c}$ Also at Novosibirsk State University, Novosibirsk, Russia\\
$^{d}$ Also at TRIUMF, Vancouver BC, Canada\\
$^{e}$ Also at Department of Physics {\&} Astronomy, University of Louisville, Louisville, KY, United States of America\\
% ...
$^{*}$ Deceased
\end{flushleft}


\end{document}